\documentclass[12pt]{iopart}
\usepackage{iopams,graphicx}

\usepackage{color}
\newcommand{\hlA}[1]{{\color{black}{#1}}} 

\begin{document}

\title[Spectrum and Franck-Condon factors of interacting suspended SWCNTs]{Spectrum and Franck-Condon factors of interacting suspended single-wall carbon nanotubes}

\author{Andrea Donarini, Abdullah Yar and Milena Grifoni}

\address{Institute of Theoretical Physics, University of Regensburg, D-93040
Regensburg, Germany}

\ead{andrea.donarini@physik.uni-r.de}
\begin{abstract}

A low energy theory of suspended carbon nanotube quantum dots in weak tunnelling coupling with metallic leads is presented. The focus is put on the dependence of the spectrum and the Franck-Condon factors on the geometry of the junction including several vibronic modes. The relative size and the relative position of the dot and its associated vibrons strongly influence the electromechanical properties of the system. A detailed analysis of the complete parameters space reveals different regimes: in the short vibron regime the tunnelling of an electron into the nanotube generates a plasmon-vibron excitation while in the long vibron regime polaron excitations dominate the scenario. The small, position dependent Franck-Condon couplings of the small vibron regime convert into uniform, large couplings in the long vibron regime. Selection rules for the excitations of the different plasmon-vibron modes via electronic tunnelling events are also derived.

\end{abstract}

\submitto{\NJP}

\maketitle

\section{Introduction}

The nanoelectromechanical systems (NEMS) are characterized by a peculiar interplay between electronic and mechanical degrees of freedom ~\cite{ekinci2005}. Suspended carbon nanotubes constitute a particularly interesting realization of NEMS due to their remarkable electronic and vibronic properties ~\cite{saito1998cnt,suzuuraPRB2002,sazonovanature2004}. \hlA{NEMS can be realized, though, in a variety of different flavors, with single molecule junctions~\cite{park2002,yu2004,qiu2004,park2000,smit2002}, suspended and laterally confined two dimensional electron gases \cite{weig2004}, silicon \cite{blick2010}, or suspended graphene \cite{bachtold2011}}. Interesting vibrational effects in electronic transport have been observed in several recent experiments on suspended single wall carbon nanotube (SWCNT) quantum dots~\cite{sapmaz2006,huettel2009,leturcq09}. These experimental works have triggered several attempts ~\cite{flensbergnjp2006,leturcq09,izumida2005,CavalierePRB(R)2010} to theoretically explain some characteristic features of the measured stability diagrams ({\it i.e.}, of the differential conductance in a bias voltage-gate voltage colour map).
In particular, the height of the conductance peaks associated with the vibronic resonances is in quantitative agreement ~\cite{sapmaz2006} with predictions of a simple Franck-Condon model for a single electronic level coupled to a harmonic mode (the so called Anderson-Holstein model) ~\cite{boseEL2001,braigPRB2003,mitraPRB2004,kochPRL2005,jkochPRB2006}. Nevertheless, the size of the electron phonon couplings required to fit the experimental data has remained essentially impossible to achieve with a microscopic theory \cite{flensbergnjp2006,izumida2005}, without introducing large screening effects \cite{CavalierePRB(R)2010}.
Moreover, the experimental data present negative differential conductance features which go beyond the capability of the simple Anderson-Holstein model. Different extensions of this model \cite{SchultzPRB2010, ZazunovPRB2006, ShenPRB2007,YarPRB2011} including asymmetric coupling or multiple electronic levels  have been proposed to explain NDCs. In a recent work~\cite{CavalierePRB(R)2010}, they have been attributed to a spatial dependent Franck-Condon factor, as it naturally occurs in a clamped nanotube, combined with the assumption of a vibron mode being mostly localized near one of the two dot ends.

Convinced of the fundamental importance of the geometrical configuration of the junction on the transport characteristics of a suspended SWCNT, we improve and extend here the work presented in \cite{CavalierePRB(R)2010}. Specifically, we calculate the spectrum and the tunnelling matrix elements  over the entire parameters space obtained varying the relative length and relative position of the vibron with respect to the quantum dot also including the effect of higher vibronic modes. The treatment of a wide parameters space is relevant since it allows for a unified picture of different results presented in the literature \cite{izumida2005,CavalierePRB(R)2010}. Also the inclusion of the higher vibronic modes seemed to us a necessary extension for two reasons: there is no real energy separation between the different vibrational modes since the frequency of the $n$th vibronic mode is just and $n$th multiple of the fundamental frequency $\omega$ of the lowest one; furthermore, the very same dispersion relation (linear with respect to the mode number $n$) implies the presence of several degenerate vibronic configurations for the system, a necessary condition for interference triggered NDC features in the stability diagrams of nanojunctions in the single electron transistor set-up \cite{begemannPRB2008,donarininanolett2009,darau2009,donariniPRB2010,schultz2010}.

The spectrum is obtained via the exact diagonalization of the system Hamiltonian with the mechanical degrees of freedom being coupled both to the total charge and the plasmons of the nanotube. As a result, at low  energies the system is described by a set of displaced vibron-plasmon excitations and the tunnelling matrix elements reduce to the product of Franck-Condon factors, one for each vibron-plasmon mode. Importantly, the Franck-Condon couplings are different for the different modes and depend crucially on the geometry of the system. A detailed analysis of the complete parameters space reveals different regimes: in the short vibron regime the tunnelling of an electron into the nanotube generates a plasmon-vibron excitation while in the long vibron regime the polaron excitation dominates the scenario. The small, position dependent Franck-Condon couplings of the small vibron regime convert into uniform, large couplings in the plasmonic case.

\begin{figure}[h!]
\begin{center}
  \includegraphics[width=\textwidth]{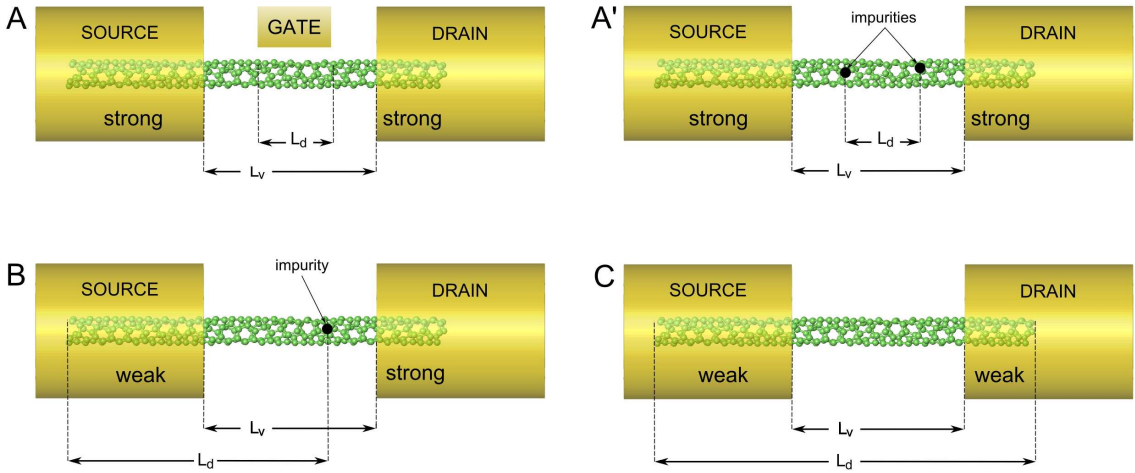}
  \caption{\hlA{Different realizations of a nanojunction with a suspended SWCNT. The lengths $L_{\rm d}$ of the quantum dot and $L_{\rm v}$ of the vibrons are also indicated together with their position. The length and position of the vibrons is assumed to coincide with the suspended part of the tube. The length and position of the dot depend instead on many factors like for example the weak or strong hybridization of the SWCNT and the metallic leads and the presence of impurities or of side gates. The labelling of the different configurations is given according to the general one used in figure \ref{fig:Phase_space}}.}
  \label{fig:Set-up}
\end{center}
\end{figure}

The paper is organized as follows: In section 2 the model Hamiltonian of a suspended SWCNT coupled to several stretching modes is introduced. Particular emphasis is given to the dependence of the electron-vibron coupling on the geometry of the system, see figure \ref{fig:Set-up}. A detailed analysis of the coupling constants as a function of the geometrical parameters is performed.
A set of canonical transformation including a polaron transformation is employed in section 3 to obtain the spectrum of the SWCNT in the presence of electron-electron and electron-vibron interactions. Both analytical results for limiting cases and general numerical findings (see figure \ref{fig:DeltaE}) on the entire parameters space are discussed.
As known from the theory of Franck-Condon blockade in the simplest Anderson-Holstein model~\cite{braigPRB2003,kochPRL2005}, the polaron transformation also crucially affects the tunnelling Hamiltonian describing the coupling to the source and drain leads. In section 4  an analytical expression of the tunnelling matrix elements is derived. A detailed analysis of the associated Franck-Condon couplings is performed, revealing different regimes and selection rules for the tunnelling processes depending on the geometrical configurations. Conclusions are drawn in section 5.

\section{Low-energy Hamiltonian of suspended SWCNTs}

The low energy spectrum of finite size, interacting metallic SWCNTs has been discussed in \cite{oregPRL2000} within a mean field approach and in \cite{LeoPRB2006,leoepjb2008} within a bosonization framework going beyond the mean field results. Bosonization is also the natural approach to include the effects of the coupling to the longitudinal stretching modes~\cite{izumida2005,CavalierePRB(R)2010}. Here, following \cite{izumida2005,CavalierePRB(R)2010,LeoPRB2006}, we derive and discuss the spectrum and many body states of suspended metallic SWCNTs. Particular emphasis will be given to the dependence of the electron-vibron coupling on the geometrical configuration of the system. An analytical expression of the electron-vibron coupling constants in terms of the relevant geometrical parameters is derived and plays a crucial role in the analysis of the spectrum and the matrix elements conducted in the following sections.

We thus consider a Hamiltonian of the form
\begin{equation}\label{eq:H_sys}
\hat{H}_{\rm sys} = \hat{H}_{\rm 0}+\hat{V}_{\rm ee}+\hat{H}_{\rm v}+\hat{H}_{\rm ev},
\end{equation}
where $\hat{H}_0$ is the noninteracting Hamiltonian of a finite size, metallic SWCNT, $\hat{V}_{\rm ee}$ describes the electron-electron interaction, $\hat{H}_{\rm v}$ is associated with the longitudinal stretching modes while $\hat{H}_{\rm ev}$ describes the electron-vibron coupling.

\subsection{Metallic nanotubes at low energies}
 Exemplarily we shall perform the quantitative analysis for armchair SWCNTs. The extension to arbitrary chiralities, though, does not change neither the essence of the calculations nor the main results presented here. In armchair SWCNTs at low energies and under periodic boundary conditions only the gapless energy subbands with linear dispersion touching at the Fermi points $F=\pm K_0 \hat{e}_x$ (also denoted Dirac points), where $\hat{e}_x$ is directed along the tube axis, are relevant~\cite{EggerPRL1997,KanePRL1997}. Imposing open boundary conditions along the tube length $L_{\rm d}$, the eigenfunctions of the noninteracting Hamiltonian $\hat{H}_0$ are the standing waves~\cite{LeoPRB2006,leoepjb2008}
\begin{equation}\label{eq:Standing_waves}
\varphi_{r\kappa}\left( \vec{r}\right)=\frac{1}{\sqrt{2}}\sum_{F}\mathrm{sgn}(F)e^{i\mathrm{sgn}(F)\kappa x}\sum_p f_{pr}\varphi_{pF}\left( \vec{r}\right) ,
\end{equation}
where $\varphi_{pF}\left( \vec{r}\right)$ describes fast oscillating Bloch waves on the graphene sublattice $p=\pm$ at the Fermi point $F$. The branch index $r=\pm$ accounts for left (+) and right (-) moving electrons. For armchair SWCNTs it holds $f_{pr}=\frac{1}{\sqrt{2}}$ if $p=+$ and $f_{pr}=-\frac{r}{\sqrt{2}}$ if $p=-$.
The parameter
\begin{equation}\label{eq:Momenta}
\kappa=\frac{\pi}{L_{\rm d}}\left(n_\kappa+\Delta \right),\quad n_\kappa\in Z,\quad |\Delta|<\frac{1}{2},
\end{equation}
measures the wave number with respect to the Fermi points $K_0$, while $\Delta$ has to be introduced if there is no integer $n$ with $K_0=\frac{\pi n}{L_{\rm d}}$. Including the spin degree of freedom $\sigma$ the Hamiltonian $\hat{H}_0$ therefore reads
\begin{equation}\label{eq:Kinetic_H}
\hat{H}_{0}=\hbar v_F\sum_{r\sigma}r\sum_\kappa \kappa \hat{c}^\dagger _{r\sigma\kappa }\hat{c}_{r\sigma\kappa },
\end{equation}
where $v_F\approx 8.1\times 10^5 {\rm m/s}$ is the Fermi velocity. Thus the level spacing of the noninteracting system is given by $\varepsilon_0=\hbar v_F\frac{\pi}{L_{\rm d}}$ while $\varepsilon_\Delta=\varepsilon_0\Delta$ denotes the energy mismatch between the $r=\pm$ branches. Moreover, the operator $\hat{c}_{r\sigma\kappa }$ annihilates an electron in the state $|\varphi_{r\kappa}\rangle|\sigma\rangle$. In turn, the electron field operator is expressed in terms of the wave functions $\varphi_{r\kappa}\left( \vec{r}\right)$ as
\begin{equation}\label{eq:electronoperator}
\hat{\Psi}_\sigma\left( \vec{r}\right)=\sum_{r\kappa} \varphi_{r\kappa}\left( \vec{r}\right) \hat{c}_{r\sigma\kappa }.
\end{equation}
The electron-electron interaction assumes the standard form
\begin{equation}\label{eq:coulombinteraction}
\hat{V}_{\rm ee}=\frac{1}{2}
\sum_{\sigma,\sigma' }
\int {\rm d}{\vec{r}}\int {\rm d}{\vec{r}\,'}
\hat{\Psi}^\dagger_{\sigma}(\vec{r})\hat{\Psi}^\dagger_{\sigma'}(\vec{r}\,')
U(\vec{r}-\vec{r}\,')
\hat{\Psi}_{\sigma'}(\vec{r}\,')\hat{\Psi}_{\sigma}(\vec{r}),
\end{equation}
where for the actual calculations we model $U(\vec{r}-\vec{r}\,')$ by the so called Ohno potential~\cite{leoepjb2008}
\begin{equation}\label{eq:Ohno_potential}
U(\vec{r}-\vec{r}\,')=
U_0
\left[
1+\left(
\frac{U_0\epsilon|\vec{r}-\vec{r}\,'|}{\alpha}
\right)^2
\right]^{-1/2},
\end{equation}
where a reasonable choice of the onsite energy is~\cite{leoepjb2008} $U_0=15\,{\rm eV}$, the dielectric constant is $\epsilon\approx 1.4-2.4$ and ${\alpha} = 14.397\,{\rm eV}{\rm \AA}$.

The Coulomb interaction causes Umklapp, backward and forward scattering processes among the electrons. Away from half filling it is reasonable to neglect Umklapp scattering. We also disregard backscattering processes, which is a valid approximation for nanotubes with not too small radii \cite{leoepjb2008}. The forward scattering processes can be fully included within a Tomonaga Luttinger (TL) model for SWCNTs \cite{izumida2005,EggerPRL1997} yielding the TL Hamiltonian:
\begin{equation}\label{eq:TL_Hamiltonian}
\hat{H}_{\rm 0} + \hat{V}_{\rm ee} \approx \hat{H}_{\rm TL} = \hat{H}_{\rm N} + \sum_j \hat{H}_j,
\end{equation}
where $\hat{H}_{\rm N}$ describes the fermionic configuration of the nanotube and $\hat{H}_j$ represents the bosonic excitations with the index $j = c+,s+,c-,s-$ labeling the four excitation sectors for total charge, total spin and relative (with respect to the two electronic subbands) charge and relative spin, respectively. The fermionic component of
(\ref{eq:TL_Hamiltonian}) can be casted into the form:
\begin{equation}\label{eq:H_N}
\hat{H}_{\rm N}
 = \frac{\varepsilon_0}{4}\sum_j \frac{\hat{N}_j^2}{2} + \varepsilon_\Delta \hat{N}_{c-} + E_{\rm c}\frac{\hat{N}_{c+}^2}{2}
\end{equation}
where the particle number operators for the different charge and spin sectors are defined by
$\hat{N}_{c+} = \sum_{r\sigma}\hat{N}_{r\sigma}$, $\hat{N}_{c-} = \sum_{r\sigma}
{\rm sgn}(r)\hat{N}_{r\sigma}$, $\hat{N}_{s+} = \sum_{r\sigma}{\rm sgn}(\sigma)\hat{N}_{r\sigma}$ and $\hat{N}_{s-} = \sum_{r\sigma}
{\rm sgn}(r\sigma)\hat{N}_{r\sigma}$, and the operator $\hat{N}_{r\sigma}$ counts the particles with spin $\sigma$ and pseudospin $r$. The electron-electron interaction is parametrized, in the fermionic part of the Hamiltonian, by $E_{\rm c}$, {\it i.e.} the charging energy of the SWCNT quantum dot. Finally, $H_j$ describes the bosonic excitation of the sector $j$. \hlA{In the long wavelength limit, it reads}:
\begin{equation}\label{eq:H_j}
\hat{H}_j = \frac{\varepsilon_0}{g_j}
\sum_{n \geq 1} n\,\hat{b}^\dagger_{j,n}\hat{b}_{j,n},
\end{equation}
where the sum runs over the mode number $n$.  Due to the Coulomb interaction the factor $g_j<1$ for the sector $c+$ while $g_j = 1$ for the other cases. For unscreened interaction $g_{c+} \approx 0.2$ \cite{leoepjb2008,EggerPRL1997}.

\subsection{The electron-vibron Hamiltonian}
The low-energy vibrational excitations of the nanotube can be described in terms of low-energy acoustic modes~\cite{suzuuraPRB2002,marianiPRB2009,martino2003}.
These modes are coupled to the electronic degrees of freedom either via a deformation potential (associated with local variations in area) or to bond lengths modifications.
The latter coupling mechanism has a coupling constant one order of magnitude smaller than the one associated with the deformation potential~\cite{suzuuraPRB2002}. Hence the twisting modes which involve pure shear and thus a modification of the bond length can be neglected. Likewise the bending and breathing modes, though coupled via the dominant deformation potential, do not play a significant role~\cite{suzuuraPRB2002,marianiPRB2009}. In fact, the bending modes only couple quadratically to the electronic degrees of freedom, while the breathing modes lie too high in energy to be excited in low-bias transport experiments. Thus, in doubly clamped SWCNTs, the stretching modes only can be retained, in agreement with experimental conclusions~\cite{sapmaz2006}.\\ Following Ref.~\cite{izumida2005} the stretching mode Hamiltonian is expressed in a continuum model as
\begin{equation}\label{eq:vibronhamiltonian}
\hat{H}_{\rm v}=\frac{1}{2}
\int_{x_{\rm v}-\frac{L_{\rm v}}{2}}^{x_{\rm v}+\frac{L_{\rm v}}{2}} \rmd x \left[\frac{1}{\zeta}\hat{P}^2(x)+\zeta v^2_{st}\left(\partial_{x}\hat{u}(x)\right)^2 \right],
\end{equation}
where $\zeta=2\pi RM$, with $R$ the tube radius, $M$ is the carbon mass per unit area and $v_{\rm st}$ is the velocity of the longitudinal stretching mode. Moreover, $x_{\rm v}$ and $L_{\rm v}$ are the position of the center and the length of the vibron, respectively. Typical SWCNT parameters are: $v_{\rm st}=2.4\times 10^4 {\rm m/s}$, $M=3.80\times 10^{-7}{\rm kg/m^2}$.

Notice that the positions and the lengths of the dot and of the vibron do not necessarily coincide. The length of the vibron ($L_{\rm v}$) is readily estimated as the distance between the electrodes which clamp the nanotube and it is defined as the length of the free standing portion of the tube. Instead, the relation between the size $L_{\rm d}$ of the quantum dot and the geometrical properties of the junction is much more complex. The best way to estimate $L_{\rm d}$ is to extract it from transport measurements which give the mean level spacing and the charging energy of the system. The position of the center of the dot $x_{\rm d}$ can be taken as a free parameter.

\hlA{In figure \ref{fig:Set-up} we sketch four possible physical realizations of different configurations. In the panels A and A' the dot lies inside the vibrating part of the tube. The confinement is obtained by a side gate (A) or by impurities located on the tube (A'), while the rest of the tube is electrically absorbed into the leads due to the strong tube-lead hybridization (extended lead configuration). In panel C the dot coincides with the entire tube length due to the weak hybridization between the SWCNT and the metallic leads and fully contains the vibrating fraction of the tube. Finally, a somehow mixed scenario is illustrated in panel B.}

The electron-vibron coupling Hamiltonian reads:
\begin{equation}\label{eq:ev_int}
\hat{H}_{\rm ev}=\int \rmd {\vec{r}} \hat{\rho} (\vec{r}) \hat{V}(\vec{r}),
\end{equation}
where  $\hat{\rho}(\vec{r})=\sum_\sigma\hat{\Psi}^\dagger_{\sigma}(\vec{r})\hat{\Psi}_\sigma (\vec{r})$ is the electron-density and $\hat{V}(\vec{r})=g\partial_{x}\hat{u}(x)$ is the deformation potential for the stretching vibron mode. The coupling constant $g$ is estimated to be~\cite{suzuuraPRB2002} $g\approx 20-30{\rm eV}$.
The displacement and momentum field operators read~\cite{landauTE1970}
\begin{equation}\label{eq:uP_operators}
\eqalign{
\hat{u}(x)=\sqrt{\frac{\hbar}{\zeta v_{\rm st}L_{\rm v}}}
\sum_{m\geq 1}\sin\left[k_m \left(x-x_{\rm v}+\frac{L_{\rm v}}{2}\right)\right]\frac{1}{\sqrt{k_m}}
\left( \hat{a}^\dagger_m + \hat{a}_m \right),\\
\hat{P}(x)=i\sqrt{\frac{\hbar\zeta v_{\rm st}}{L_{\rm v}}}
\sum_{m \geq 1}
\sin\left[k_m \left(x-x_{\rm v}+\frac{L_{\rm v}}{2}\right)\right]\sqrt{k_m}
\left( \hat{a}^\dagger_m - \hat{a}_m \right),}
\end{equation}
with  $k_m= m\pi/L_{\rm v}$ the wave number. Here $\hat{a}_{m}(\hat{a}^\dagger_{m})$ are the annihilation (creation) operators associated with the $m$th vibron mode obeying the commutation relation $[\hat{a}_m, \hat{a}^\dagger_{m'}]= \delta_{m,m'}$.
Using the above field operators, we obtain
\begin{equation}\label{eq:H_v}
\hat{H}_{\rm v}=\sum_{m\geq 1}E_m
\left( \hat{a}^\dagger_m \hat{a}_m+\frac{1}{2} \right)  ,
\end{equation}
with  $E_m=m\hbar v_{\rm st}\pi/L_{\rm v}\equiv m\hbar\omega$.
Similarly (\ref{eq:ev_int}), integrated over radial and azimuthal coordinates, becomes

\begin{equation}\label{eq:ev_int2}
 \fl \hat{H}_{\rm ev}= g\sum_{m\geq 1} \left(\frac{\hbar k_{m}}{\zeta v_{\rm st}L_{\rm v}} \right)^{1/2}\left(\hat{a}^\dagger_{m}+\hat{a}_{m}\right)
 \int_{{\rm d}\cap{\rm v}}\!\!\! {\rm d}x\hat{\rho}_{\rm 1D}(x)\cos\left[k_{m} \left( x - x_{\rm v}+\frac{L_{\rm v}}{2}\right)\right],
\end{equation}
where the integral is calculated over the overlap of the dot and vibron region and the effective 1D density operator $\hat{\rho}_{\rm 1D}(x)$ reads, in its bosonized form \cite{LeoPRB2006}:
\begin{equation}\label{eq:rho_bosonized}
\hat{\rho}_{\rm 1D}(x) = \frac{\hat{N}_{c+}}{L_{\rm d}} + \frac{2}{\sqrt{\pi\hbar}}\partial_{x}\hat{\phi}_{c+}(x).
\end{equation}
Notice that the bosonic field $\hat{\phi}_{c+}(x)$ can be expressed in terms of the bosonic creation and annihilation operators $\hat{b}^{\dagger}_{c+,n}$ and $\hat{b}_{c+,n}$ as:
\begin{equation}\label{eq:Boson_field}
\hat{\phi}_{c+}(x) = \sqrt{\frac{\hbar g_{c+}}{L_{\rm d}}}
\sum_{n \geq 1} \sin\left[k_n\left( x-x_{\rm d}+\frac{L_{\rm d}}{2}\right)\right]\frac{1}{\sqrt{k_n}}(\hat{b}_{c+,n}^\dagger + \hat{b}_{c+,n})
\end{equation}
where $k_n = n\pi /L_{\rm d}$ and we have imposed open boundary conditions $\hat{\rho}(x_{\rm d}-L_{\rm d}/2) = \hat{\rho}(x_{\rm d}+L_{\rm d}/2) = 0$.
It is useful, for the diagonalization  procedure presented in the next subsection, to introduce the dimensionless position and momentum operators for the $n$th plasmon mode $\{\hat{X}_n,\hat{P}_n\}$ and the ones of the $m$th vibronic mode $\{\hat{x}_m,\hat{p}_m\}$
\begin{equation}\label{eq:XP_plasmon}
\eqalign{
\hat{X}_n =
\frac{\hat{b}_{c+,n}+\hat{b}^\dagger_{c+,n}}{\sqrt{2}}, \qquad
\hat{x}_m =
\frac{\hat{a}_m+\hat{a}^\dagger_m}{\sqrt{2}},\\
\hat{P}_n = \frac{\hat{b}_{c+,n}-\hat{b}^\dagger_{c+,n}}{\rmi\sqrt{2}},
\qquad
\hat{p}_m =
\frac{\hat{a}_m-\hat{a}^\dagger_m}{\rmi\sqrt{2}},\\
}
\end{equation}
which satisfy the canonical commutation relations $[\hat{X}_n,\hat{P}_{n'}] = \rmi \delta_{nn'}$ and $[\hat{x}_m,\hat{p}_{m'}] = \rmi \delta_{mm'}$. In terms of these operators the electron-vibron Hamiltonian can be written as:

\begin{equation}\label{eq:ev_int3}
\eqalign{
\hat{H}_{\rm ev} &= I \sqrt{g_{c+}} \sum_{n,m \geq 1} \sqrt{nm}K_{nm}(\lambda,\delta) 2 \hat{X}_n \hat{x}_m\\
&+ I \sum_{m \geq 1}\sqrt{m} L_m (\lambda,\delta)\sqrt{2} \hat{N}_{c+}\hat{x}_m}
\end{equation}
where
\begin{equation}\label{eq:ev_Interaction_constant}
I = g \sqrt{\frac{\hbar \pi}{\zeta v_{\rm st}L_{\rm d}^2}}
\end{equation}
is the fundamental coupling constant; it acquires the value $I = 88\, \mu {\rm eV}$ for a $(10,10)$ SWCNT with $L_{\rm d} = 1\,\mu {\rm m}$ and assuming $g = 30\,{\rm eV}$.

The \emph{geometric} part of the electron-vibron coupling is given by the dimensionless matrix
\begin{equation}\label{eq:K_definition}
\eqalign{
 K_{nm}(\lambda,\delta) =
 \frac{1}{\lambda} \int_{x_{\rm min}}^{x_{\rm max}} \!\!\rmd x
  &\left\{\cos\left[\pi x \left(n+\frac{m}{\lambda}\right) -\frac{m\pi}{\lambda}\left(\delta+\frac{1-\lambda}{2}\right)\right]
  +\right.\\
  &\phantom{,,}\left.\cos\left[\pi x \left(n-\frac{m}{\lambda}\right) +\frac{m\pi}{\lambda}\left(\delta+\frac{1-\lambda}{2}\right)\right]
  \right\}
  }
\end{equation}
for the plasmon-vibron component and by the vector
\begin{equation}\label{eq:L_definition}
L_m(\lambda,\delta) = \frac{1}{\lambda}
\int_{x_{\rm min}}^{x_{\rm max}}
\rmd x \cos\left[\frac{m\pi}{\lambda}\left(x-\delta -\frac{1-\lambda}{2}\right)\right]
\end{equation}
for the charge-vibron component. The integration limits
\begin{equation}\label{eq:int_limits}
\eqalign{
x_{\rm min} = \max[0,\delta + (1-\lambda)/2],\\
x_{\rm max} = \min[1,\delta + (1+\lambda)/2]
}
\end{equation}
ensure that the integral extends only on the overlap regions of the dot and vibron. As one appreciates from (\ref{eq:ev_int3})-(\ref{eq:L_definition}), for a fixed $L_{\rm d}$  the electron-vibron Hamiltonian is completely determined by the relative position of the centers of the dot and the vibron $\delta = (x_{\rm v}-x_{\rm d})/L_{\rm d}$, and the ratio between the length of the vibron and of the dot $\lambda = L_{\rm v}/L_{\rm d}$.

Importantly, $\hat{H}_{\rm ev}$ reveals that the electron-vibron interaction only involves the position operator $\hat{x}_m$ of the $m$-vibron mode, the position operator $\hat{X}_n$ of the $n$-th charged plasmon mode and the total electron number $\hat{N}_{c+}$.
Moreover, the important energy scales involved in the electron-vibron dynamics are the lowest vibron energy $\hbar\omega$, the lowest charged plasmon energy $\varepsilon_0/g_{c+}$ and the fundamental coupling constant $I$: their values are $0.050 meV$, $8.293 meV$ and $0.088 meV$, respectively, for a (10,10) SWCNT with $L_{\rm d} = L_{\rm v} = 1\mu m$ and the other parameters as the ones given in figure \ref{fig:DeltaE}. Excluding the extreme short vibron regime $\lambda \leq 1/100$ and the strong screening, we can conclude that $\hbar \omega, I \ll \varepsilon_0/g_{c+}$ thus implying a clear separation of the vibron and plasmon energy scales. Albeit these two degrees of freedom are consequently characterized, for an isolated system, by an essentially independent dynamics, the tunnelling event can be substantially influenced by the mechanical motion of the nanotube under certain geometrical conditions as will be discussed later.

\subsection{Plasmon-vibron and charge-vibron couplings}

 The energy spectrum and the Franck-Condon couplings strongly depend on the geometry of the junction via the coupling constants $K_{nm}$ and $L_m$. The detailed analysis of these coupling constants is thus the natural starting point to understand the presence of geometrical dependent trends and selection rules in the tunnelling processes of a suspended SWCNT.

\begin{figure}[h!]
\begin{center}
  \includegraphics[width=0.70\textwidth]{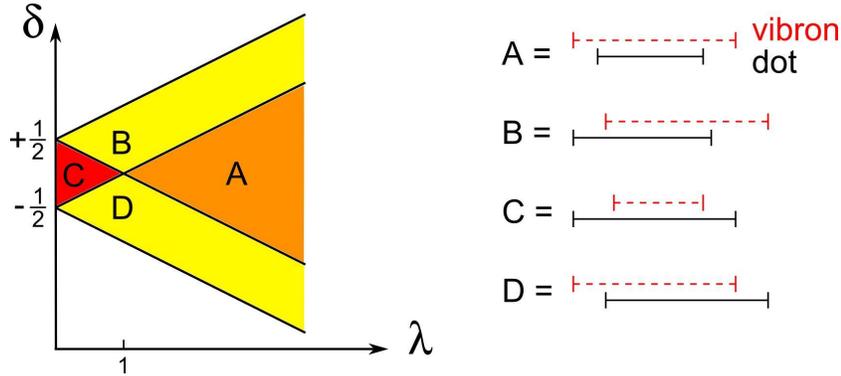}
  \caption{Parameters space of the geometrical configurations of the electromechanical nanojunction. The relevant dimensionless parameters are the length ratio $\lambda = L_{\rm v}/L_{\rm d}$ and the relative position of the centers $\delta = (x_{\rm v}-x_{\rm d})/L_{\rm d}$. Four qualitatively different regions are identified in the parameters space and schematically shown on the right.}
  \label{fig:Phase_space}
\end{center}
\end{figure}

The geometrical parameters space $\{\lambda,\delta\}$ is divided into four regions by the different conditions imposed by the  integration limits $x_{\rm min}$ and $x_{\rm max}$ explicitly given in (\ref{eq:int_limits}). In figure \ref{fig:Phase_space} we define these regions and give a schematic representation of the corresponding geometrical configuration.

The function $K_{nm}$ has the following explicit form in the four regions:

\begin{equation} \label{eq:K_explicit}
\fl
\eqalign{
K_{nm}^{(A)}(\lambda,\delta) =
-\frac{2m}{\pi(\lambda^2n^2-m^2)}
\left[
(-1)^n
\sin\left(m\pi\frac{1-2\delta+\lambda}{2\lambda}\right)+
\sin\left(m\pi\frac{1+ 2\delta -\lambda}{2\lambda}\right)
\right]\\
K_{nm}^{(B)}(\lambda,\delta) =
-\frac{2}{\pi(\lambda^2n^2-m^2)}
\left[
(-1)^n
m\sin\left(m\pi\frac{1-2\delta+\lambda}{2\lambda}\right)+
\lambda n\sin\left(\lambda n\pi\frac{1+ 2\delta-\lambda }{2\lambda}\right)
\right]\\
K_{nm}^{(C)}(\lambda,\delta) =
\frac{2\lambda n}{\pi(\lambda^2n^2-m^2)}
\left[
(-1)^m
\sin\left(\lambda n\pi\frac{\lambda+2\delta+1}{2 \lambda}\right)+
\sin\left(\lambda n\pi\frac{\lambda - 2\delta -1}{2 \lambda}\right)
\right]\\
K_{nm}^{(D)}(\lambda,\delta) =
\frac{2}{\pi(\lambda^2n^2-m^2)}
\left[
(-1)^m
\lambda n \sin\left( \lambda n\pi\frac{\lambda+2\delta+1}{2\lambda}\right)+
m \sin\left(m\pi\frac{\lambda+ 2\delta -1}{2\lambda}\right)
\right]\\
}
\end{equation}
Some symmetry relations for the function $K_{nm}$ can be readily obtained from (\ref{eq:K_explicit}):
\begin{equation}\label{eq:symmetry_K}
\eqalign{
K_{nm}(\lambda,\delta) = 1/\lambda\, K_{mn}(1/\lambda,-\delta/\lambda),\\
K_{nm}(\lambda,\delta) = (-1)^{n+m}K_{nm}(\lambda,-\delta).\\
}
\end{equation}
The first equation in (\ref{eq:symmetry_K}) quantifies the connection between the behaviour of $K_{nm}(\lambda,\delta)$ at small and large values of $\lambda$: the roles of the vibron and of the plasmon are simply exchanged in the plasmon-vibron Hamiltonian if we invert the ratio of their lengths. The second equation in (\ref{eq:symmetry_K})
states instead that if we invert the relative position of the vibron and the dot, the plasmon-vibron Hamiltonian acquires at most a minus sign, depending on the parity of the vibronic and plasmonic modes. The case considered in \cite{izumida2005} corresponds to the point $\lambda = 1$, $\delta = 0$ of the parameters space where the following limit holds:
\begin{equation}\label{eq:K_limits_1}
\lim_{\lambda \to 1}K_{nm}(\lambda,0) = \delta_{nm},\\
\end{equation}
and each vibronic mode is only coupled to the plasmonic mode of the same order $n = m$. In \emph{all} other regions of the parameters space the coupling is not diagonal and the dynamics of each vibronic mode is influenced by all plasmonic modes and vice versa, making the system quite intricate. Nevertheless, from a detailed analysis of the $K$ function, one can estimate which modes are more relevant in the low energy limit.

The function $K$ has an upper bound $K < 2$, as it can be easily proven from its definition (\ref{eq:K_definition}) by considering that the distance between the integration limits is at maximum $\lambda$. Thus, $K_{nm}$ does not diverge for $\lambda \to m/n$ as one could expect from a  first sight. Instead, its maximum can be estimated by calculating the limit $\lambda \to m/n$. One obtains:

\begin{equation}\label{eq:K_limits_2}
\lim_{\lambda \to \frac{m}{n}} K_{mn}(\lambda,\delta) =
\cases{
\frac{n}{m}\cos\left[\frac{\pi}{2}(n-m-2n\delta)\right]&  for  $n<m$,\\
\cos\left[\frac{\pi}{2}(n-m-2n\delta)\right] & for $n>m,$
}
\end{equation}
where the first and the second case are calculated in region A and C, respectively (see figure \ref{fig:Phase_space}). The absolute value $|K_{nm}|$ of the coupling exhibits $|m-n|+1$ local maxima as a function of the relative displacement $\delta$, separated by nodes in which the $m$-th vibronic mode is decoupled from the $n$-th plasmonic one. Notice that in the limit $\lambda \to m/n$ the wavelength of the $m$-th vibronic mode coincides with the one of the $n$-th plasmonic one, thus giving a physical interpretation to the resonance. One concludes that each geometrical configuration optimizes the coupling between specific plasmonic and vibronic modes. Moreover, the coupling between low vibronic modes and higher plasmonic ones is reached for short vibrons and is more efficient than the coupling of a low plasmonic mode with higher vibronic ones obtained, instead, for large vibrons.

\begin{figure}[h!]
\begin{center}
  \includegraphics[width=0.7\textwidth]{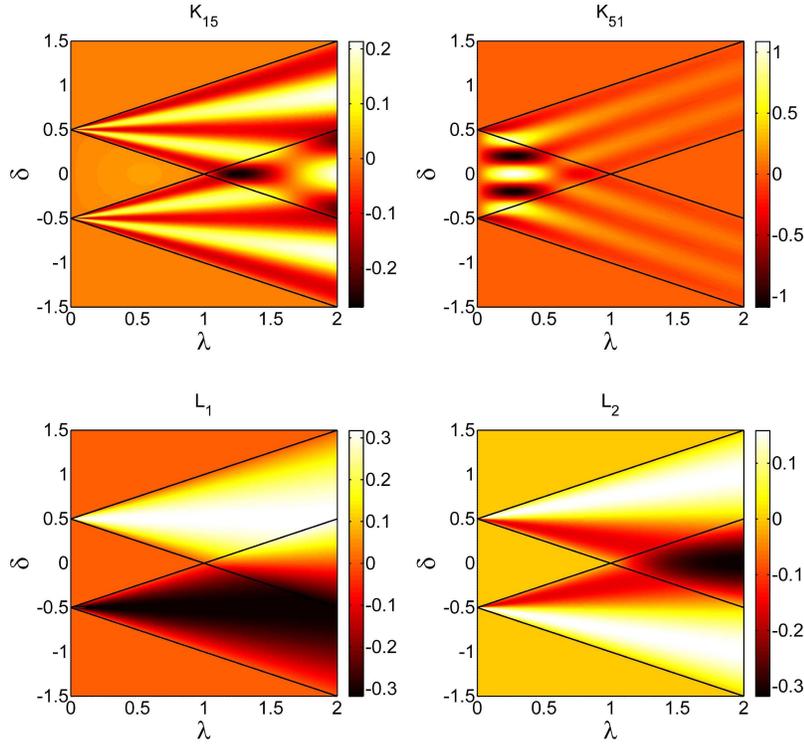}
  \caption{The plasmon-vibron $K_{nm}$ and the charge-vibron $L_m$ coupling constants are plotted in the geometrical parameters space. Top row: On the left (right) the coupling between the first (fifth) plasmonic and the fifth (first) vibronic modes. Bottom row: Examples of the charge-vibron coupling $L_m$ are given for the first (left) and the second (right) vibronic mode. Black solid lines indicate in all figures the borders of the regions A, B, C, D indicated in figure (\ref{fig:Phase_space}) and explained in the text.}
  \label{fig:KL_functions}
  \end{center}
\end{figure}

Another interesting regime can be reached in the small vibron region  when the center of the vibron lies in the vicinity of the border of the dot. Let us consider for this reason the function $K_{nm}$ in the region $B$ and with $\lambda \ll m/n$. The following relation holds:
\begin{equation}\label{eq:K_limits_3}
K_{nm}\left(\lambda,\frac{1}{2}+\alpha \lambda\right) = \frac{2}{\pi m}(-1)^n\sin\left[m\pi\left(\frac{1}{2}-\alpha\right)\right]
\end{equation}
where $|\alpha|<1/2$. The absolute value $|K_{nm}|$  of the coupling exhibits $m$ local maxima as a function of $\alpha$ in the region B which are independent of the plasmonic mode $n$. This specific configuration has been chosen in \cite{CavalierePRB(R)2010} to describe a system in which the renormalization of the lowest vibronic mode due to the coupling to the plasmons produces a strongly inhomogeneous Franck Condon coupling in the tunnelling matrix elements to the carbon nanotube. In order to illustrate the arguments presented so far, we plot in figure (\ref{fig:KL_functions}) the plasmon-vibron  couplings  $K_{15}$ and $K_{51}$ as a function of $\lambda$ and $\delta$. Clearly visible are the maxima of the coupling close to $\lambda = m/n$ and the fan shape structure  of the coupling close to the points $\{0 , \pm 1/2\}$ more visible in the case $K_{15}$ due to the conditions given above.

The second line of the electron-vibron Hamiltonian (\ref{eq:ev_int3}) describes the charge-vibron interaction and is proportional to the function $L_m$ defined in (\ref{eq:L_definition}). Also the coupling $L_m$, is defined on the parameters space $\{\lambda,\delta\}$ by different functions in the four different regions already introduced for $K_{nm}$:

\begin{equation}\label{eq:L_explicit}
\eqalign{
L^{(A)}_m(\lambda,\delta) = \frac{1}{m\pi}
\left[
\sin\left(m\pi\frac{1-2\delta+\lambda}{2\lambda}\right) + \sin\left(m\pi\frac{1+2\delta-\lambda}{2\lambda}\right)
\right],\\
L^{(B)}_m(\lambda,\delta) = \frac{1}{m\pi}
\sin\left(m\pi\frac{1-2\delta+\lambda}{2\lambda}\right),\\
L^{(C)}_m(\lambda,\delta) = 0,\\
L^{(D)}_m(\lambda,\delta) = \frac{1}{m\pi}
\sin\left(m\pi\frac{1+2\delta-\lambda}{2\lambda}\right).
}
\end{equation}
A symmetry relation can also be derived for this coupling, namely:
\begin{equation}\label{eq:symmetry _L}
L_m(\lambda,\delta) = (-1)^m L_m(\lambda,-\delta).
\end{equation}
The function $L_m$ vanishes identically in the region $C$ thus implying no charge-vibron coupling for systems in which the vibron is entirely contained inside the dot. The finite local coupling is in fact averaged away by the sinusoidal shape of the vibron. The form of $L_m$ in the region B is readily understood with the parametrization:
\begin{equation}\label{eq:L_limits1}
L_m\left(\lambda,\pm\frac{1}{2}+\alpha\lambda \right) = \frac{1}{m\pi}\sin\left[m\pi\left(\frac{1}{2}-\alpha\right)\right]
\end{equation}
with $|\alpha|´\leq 1/2$. As shown in figure \ref{fig:KL_functions}, $|L_m|$ has indeed in the small vibron limit a fan shape with $m$ maxima of magnitude $1/(m\pi)$ separated by $m-1$ nodes. Thus, the charge-vibron coupling decreases for higher vibron modes and is also very sensitive to the geometry of the system. The geometry of the system even introduces selection rules: for example for a system with $\lambda = 0.1$ and $\delta = 1/2,\, (\alpha = 0)$ only odd modes ($m = 2a - 1,\, a \in \mathbb{N}^+$) exhibit a charge-vibron coupling.

The maximum charge-vibron coupling for the mode $m$ is reached when $|L_m| = 2/(m\pi)$ and is obtained only for vibrons larger than the dot (A region, $\lambda > 1$) and centered with respect to it ($\delta = 0$). Only even vibronic modes couple to the charge if $\delta = 0$ and $|L_m|$ exhibits $m/2$ maxima in the positions $\lambda = m/(1+2r)$ where $0 \leq r < m/2-1,\,r \in \mathbb{N}$.

In conclusion, the electron-vibron coupling is very sensitive to the geometry of the junction, both in its plasmon-vibron and charge-vibron components. In general, for more symmetric systems ($\delta \approx 0$) the plasmon-vibron component dominates the short vibron limit ($\lambda < 1$, region C) while the large vibron limit ($\lambda > 1$, region A) is dominated by  the charge-vibron interaction. Only for a strongly asymmetric system ($\delta \approx \pm 1/2$) in the short vibron limit ($\lambda \ll m/n$) the two components can have the same strength. Moreover, in general, the strength of the coupling decreases with the vibron mode.

Yet, the position of the nodes of the functions $K_{nm}$ and $L_m$ depends on the vibron and plasmon mode numbers $n$ and $m$, generating selection rules that depend sensitively on the geometry of the system.

\section{Diagonalization and spectrum}

Because the electron-vibron coupling only involves the total charge operator $\hat{N}_{c+}$ and the plasmon excitations, the part of the system Hamiltonian which is still to be diagonalized is:

\begin{equation}\label{eq:sys_bilinear_1}
\eqalign{
\hat{H}'_{\rm sys} &=
\sum_{n \geq 1} n\frac{\hbar \Omega}{2}\left( \hat{X}_n^2 + \hat{P}_n^2\right)
+
\sum_{m \geq 1} m\frac{\hbar \omega}{2}\left( \hat{x}_m^2 + \hat{p}_m^2\right)\\
&+
I \sqrt{g_{c+}}\sum_{n,m \geq 1} \sqrt{nm} K_{nm} 2 \hat{X}_n  \hat{x}_m
+
I \sum_m \sqrt{m} L_m \sqrt{2} \hat{N}_{c+} \hat{x}_m
}
\end{equation}
where we have introduced the frequencies $\Omega = \pi v_{\rm F}/(g_{c+}L_{\rm d})$ and $\omega =  \pi v_{\rm st}/L_{\rm v}$. The exact diagonalization of the Hamiltonian (\ref{eq:sys_bilinear_1}) can be achieved in two steps: i) a set of canonical transformations eliminates the plasmon-vibron component; ii) a polaron transformation eliminates the charge-vibron component. The final result is a collection of shifted plasmon-vibron oscillators.

The first step in the diagonalization is better understood by setting the plasmon-vibron part of the Hamiltonian into a quadratic form:
\begin{equation}\label{eq:bilinear_2}
\hat{H}'_{\rm sys} =
\left(
\begin{array}{c}
\hat{\bf X}\\
\hat{\bf x}\\
\hat{\bf P}\\
\hat{\bf p}
\end{array}
\right)^{\rm T}
\left(
\begin{array}{cccc}
H_{\rm pp} & H_{\rm pv} & 0 & 0 \\
H_{\rm vp} & H_{\rm vv} & 0 & 0 \\
0 & 0 & H_{\rm pp} & 0\\
0 & 0 & 0 & H_{\rm vv}
\end{array}
\right)
\left(
\begin{array}{c}
\hat{\bf X}\\
\hat{\bf x}\\
\hat{\bf P}\\
\hat{\bf p}
\end{array}
\right)
+ \hat{H}_{\rm cv}
\end{equation}
where the components of matrix $H_{\rm M}$ defining the quadratic form are given by: $\left(H_{\rm pp}\right)_{nn'} = n \hbar \Omega /2 \,\delta_{nn'}$, $\left(H_{\rm vv}\right)_{mm'} = m \hbar \omega /2 \,\delta_{mm'}$ and $\left(H_{\rm pv}\right)_{nm} = \left(H_{\rm vp}\right)_{mn} =  I \sqrt{g_{c+}}\sqrt{nm}K_{nm}$. Moreover, we have introduced the vector of operators $\hat{\bf X} = [\hat{X}_1,\hat{X}_2,\ldots]^{\rm T}$ and analogously for $\hat{\bf x},\,\hat{\bf P}$, and $\hat{\bf p}$. Finally we have defined the charge-vibron Hamiltonian $\hat{H}_{\rm cv}$. The quadratic form in (\ref{eq:bilinear_2}) is simplified via the following set of canonical transformations: the first is the contraction

\begin{equation}\label{eq:Canonical_1}
\begin{array}{ll}
\hat{X}'_n = 1/\sqrt{n\hbar \Omega}\,\hat{X}_n, &
\hat{x}'_m = 1/\sqrt{m\hbar \omega}\,\hat{x}_m,\\
\hat{P}'_n = \sqrt{n\hbar \Omega}\,\hat{P}_n, &
\hat{p}'_m = \sqrt{m\hbar \omega}\,\hat{p}_m,
\end{array}
\end{equation}
 that transforms the momentum block of $H_{\rm M}$ into the matrix ${\bf 1}/2$. Notice that the commutation relations between position and momentum operators are conserved for each mode: $[\hat{X}'_n,\hat{P}'_{n'}] = \rmi\delta_{nn'}$ and $[\hat{x}'_n,\hat{p}'_{n'}] = \rmi\delta_{nn'}$. Afterwards we perform the rotation

\begin{equation}\label{eq:Canonical_2}
\eqalign{
\hat{\xi}'_l =
\sum_{n = 1}^{N_{\rm p}} U^T_{ln}\,\hat{X}'_n +
\sum_{m = 1}^{N_{\rm v}}U^T_{lN_{\rm p} + m}\,\hat{x}'_m,\\
\hat{\pi}'_l = 
\sum_{n = 1}^{N_{\rm p}}U^T_{ln}\,\hat{P}'_n +
\sum_{m = 1}^{N_{\rm v}}U^T_{lN_{\rm p}+m}\,\hat{p}'_m,\\
}
\end{equation}
that diagonalizes the position block of $H_{\rm M}$ written in the primed variables. We have also introduced the total number of vibron (plasmon) modes $N_{\rm v}$ ($N_{\rm p}$) This can be done without loss of generality due to the presence of physical cut-off's both for the plasmonic and vibronic mode numbers. This transformation is physically the most important since it generates the position and momentum operators $\hat{\xi}'_l$ and $\hat{\pi}'_l$ which identify $N_{\rm p} + N_{\rm v}$ mixed plasmon-vibron modes. The matrix defining the quadratic form reads, in this mixed basis:

\begin{equation}\label{eq:bilinear_3}
H_{\rm M} =
\left(
\begin{array}{c|c}
{\bf \Delta} & 0 \\
\hline
0 & {\bf 1}/2
\end{array}
\right)
\end{equation}
where ${\bf \Delta}$ is a diagonal matrix whose diagonal element $\Delta_l$ defines the energy of the plasmon-vibron mode $\hbar \omega_l = \sqrt{2 \Delta_l}$. This relation becomes clear after the last canonical transformation, the expansion
\begin{equation}\label{eq:Canonical_3}
\eqalign{
\hat{\xi}_l = \sqrt{\hbar \omega_l}\,\hat{\xi}'_l,\\
\hat{\pi}_l = 1/\sqrt{\hbar \omega_l}\,\hat{\pi}'_l,
}
\end{equation}
that brings the system Hamiltonian into the form:

\begin{equation}\label{eq:bilinear_4}
\hat{H}_{\rm sys}' = \sum_{l} \frac{\hbar \omega_l}{2}
(\hat{\xi}_l^2  + \hat{\pi}_l^2) + H_{\rm cv}.
\end{equation}
The effect of the transformations (\ref{eq:Canonical_1}), (\ref{eq:Canonical_2}) and (\ref{eq:Canonical_3}) on the charge-vibron Hamiltonian $\hat{H}_{\rm cv}$ is readily obtained:

\begin{equation}
\hat{H}_{\rm cv} = I \sqrt{2} \sum_{lm} m L_m \sqrt{\frac{ \omega}{\omega_l}}U_{N_{\rm p}+m,\,l}\,\hat{\xi}_l \hat{N}_{c+}.
\end{equation}

The presence of $\hat{H}_{\rm cv}$ requires a second step in the diagonalization procedure: the polaron transformation $\hat{\tilde{H'}}_{\rm sys} = \rme^{-\hat{S}}\hat{H}_{\rm sys}' \rme^{+\hat{S}}$ where
\begin{equation}\label{eq:S_operator}
\hat{S} = \rmi
\sqrt{2}\sum_{lm}
\frac{I}{\hbar\omega_l}
m L_m
\sqrt{\frac{\omega}{\omega_l}}U_{N_{\rm p}+m,\,l}\,\hat{\pi}_l \hat{N}_{c+}
\end{equation}
yielding
\begin{equation}\label{eq:H_sys_pol}
\hat{\tilde{H'}}_{\rm sys} = \sum_{l} \frac{\hbar \omega_l}{2}
(\hat{\xi}_l^2  + \hat{\pi}_l^2) -
\sum_l \frac{I^2}{\hbar\omega_l}
\left(\sum_m L_m \sqrt{\frac{\omega}{\omega_l}}U_{N_{\rm p}+m,\,l}
\right)^2 \hat{N}_{c+}^2.
\end{equation}

Thus, the low energy spectrum of the suspended SWCNT reads

\begin{equation}\label{eq:spectrum}
E_{\vec{N},\vec{m}} = E_{\vec{N}}
+ \sum_{l} \hbar\omega_l \left(m_{l} + \frac{1}{2}\right)
+ \sum_{n,j \neq c+} n \varepsilon_0 m_{n,j}
\end{equation}
where $\vec{N} = [N_{c+}, N_{c-}, N_{s+}, N_{s-}]$ is the vector defining the electronic configuration and $E_{\vec{N}}$ the associated energy as can be computed from (\ref{eq:H_N}) and (\ref{eq:H_sys_pol}). The vector $\vec{m}$, instead, contains the occupation numbers $m_l$ of the plasmon-vibron modes and the ones ($m_{n,j},\, j\neq c+$) of the other relative charge and spin bosonic modes.

The diagonalization procedure presented here reproduces known results in some limiting cases. In the totally symmetric case ($\delta = 0,\,\lambda = 1$) where length and center of the dot and vibrons coincide, only the coupling between plasmons and vibrons with the same number of modes is allowed ($K_{nm} = \delta_{nm}$). One obtains that the matrix to be diagonalized by the rotation (\ref{eq:Canonical_2}) is:

\begin{equation}
n^2\left(
\begin{array}{cc}
\frac{\hbar^2\Omega^2}{2} & I\hbar\sqrt{\omega\Omega g_{c+}}\\
I\hbar\sqrt{\omega\Omega g_{c+}} & \frac{\hbar^2\omega^2}{2}
\end{array}
 \right)
\end{equation}
yielding the spectrum \cite{izumida2005}:

\begin{equation}
\label{eq:Eigenvalues1}
\hbar \omega_l = \sqrt{2\Delta_l} = n\hbar\sqrt{\frac{\Omega^2 + \omega^2}{2} \pm \sqrt{\left(\frac{\Omega^2 - \omega^2}{2}\right)^2 + \frac{4g_{c+}I^2\omega\Omega}{\hbar^2} }}
\end{equation}
where $l = \{n,\alpha\}$ and $\alpha = \pm$. For this symmetric configuration there is also no polaron shift since the charge-vibron coupling vanishes identically ($L_m = 0$). Also the case considered in \cite{CavalierePRB(R)2010} of a single vibron mode is reproduced by our general theory. \hlA{Under the only assumption that $\omega \ll \Omega$ one obtains}:

\begin{equation}
\omega_1 = \omega \sqrt{1-\frac{4 I^2 g_{\rm c+}}{\hbar^2\omega \Omega}\sum_{n = 1}^\infty K_{n1}^2}
\end{equation}
\hlA{which is always real for the parameters considered in the present paper. Moreover the case of short asymmetric vibrons $(\lambda \ll 1,\delta = 1/2)$ is particularly interesting since by means of (\ref{eq:K_limits_3}) one obtains also that the lowest plasmons ($n \ll 1/\lambda$) equally contribute to soften the frequency of the lowest vibron mode.}

\begin{figure}[h!]
\begin{center}
  \includegraphics[width=0.7\textwidth]{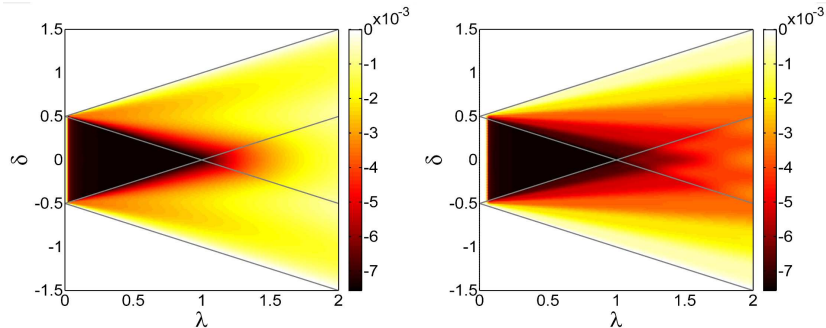}
  \caption{Relative normalization of the first (left) and fifth (right) vibron mode due to the plasmon-vibron coupling. Notice that the relative normalization is always negative, it reaches its maximum in the point $\lambda = 1,\, \delta = 0$ and is essentially  constant in the entire region C (defined in figure \ref{fig:Phase_space}). The parameters used are $L_{\rm d} = 1\,\mu{\rm m},\, R = 6.68\,{\rm \AA},\, v_{\rm F} = 8\times 10^5 \,{\rm m/s},\,v_{\rm st} = 2.4\times 10^4\,{\rm m/s},\, M = 3.8 \times 10^{-7}\,{\rm kg/m^2},\,g = 30\,{\rm eV},\,g_{c+} = 0.2.$}
  \label{fig:DeltaE}
  \end{center}
\end{figure}

 In the generic case, though, only a numerical evaluation of the spectrum is viable. In figure \ref{fig:DeltaE} we present the relative frequency shift ({\it i.e.} $(\omega_m -m\omega)/m\omega$) for the first (left panel) and the fifth (right panel) plasmon-vibron mode. The calculation is carried out for a $(10,10)$ armchair nanotube of $L_{\rm d} = 1\mu {\rm m}$. The coupling of the vibrons to the plasmons softens the vibronic modes, yielding a negative shift for every configuration. The renormalization is stronger and almost constant in the region C where the coupling between the low vibronic modes to the plasmonic ones is larger. An estimate of the maximum renormalization can be obtained by its direct calculation in the symmetric point ($\lambda = 1,\, \delta = 0$):

\begin{equation}\label{eq:renorm_sym}
\frac{\omega_m - m\omega}{m\omega} \approx -\frac{2g_{c+}I^2}{\hbar^2\omega\Omega}
\end{equation}
where we made the expansion of (\ref{eq:Eigenvalues1}) in powers of $\omega/\Omega$ and $I/(\hbar \Omega)$. Interestingly, as far as the bare vibron frequency $\omega$ and the fundamental electron-vibron coupling constant $I/\hbar$ are both much smaller than the bare plasmon frequency $\Omega$, the relative frequency normalization, if present, is independent of the mode number $m$. It is also clear that, in absence of strong screening, ($g_{c+} \approx  0.2$) the relative normalization is very moderate and does not exceed the 1 percent, independently of the geometry of the junction.

\section{Tunnelling amplitudes and Franck-Condon couplings}

So far we studied the isolated nanotube. Our interest, though, is the transport of electrons through a SWCNT in tunnelling coupling with possibly extended source and drain leads (see figure \ref{fig:Set-up}). The tunnelling Hamiltonian ${\hat{H}}_T$ is given by:
\begin{equation}\label{eq:H_tunnelling}
{\hat{H}}_{\rm T}=
\sum_{\rm \alpha=s,d}\sum_\sigma\int \rmd{\vec{r}}\left[ T_\alpha\left( \vec{r}\right){\hat{\Psi}}^\dag_\sigma\left(\vec{r} \right) \hat{\Phi}_{\sigma \alpha}\left( \vec{r}\right)+h.c. \right],
\end{equation}
where ${\hat{\Psi}}^\dag_\sigma$, see (\ref{eq:electronoperator}), and $\hat{\Phi}^\dag_{\sigma \alpha}\left( \vec{r}\right)=\sum_{\vec{q}}\phi^\ast_{\vec{q}}\left(\vec{r} \right)\hat{c}^\dag_{\vec{q}\sigma \alpha} $ are electron creation operators in the SWCNT and in the lead $\alpha$, respectively, and $T_\alpha\left( \vec{r}\right) $ describes the transparency of the tunnelling contact $\alpha$.

\hlA{The spatial dependence of the transparency $T_\alpha\left( \vec{r}\right) $ depends on the specific geometrical configuration of the junction and on the tube-lead hybridization. For the sake of simplicity we refer again to the configurations introduced in figure \ref{fig:Set-up}. In both the A and A' cases we expect $T_\alpha\left( \vec{r}\right)$ to be strongly localized at the interface between the extended lead and the dot, while in the case C the tunnelling region extends over the entire fraction of the tube which is covered by the leads. For the case B an intermediate situation is envisaged with an extended tunnelling region (weak hybridization) at the source and a localized one at the drain (strong hybridization).}

In the weak tunnelling limit the dynamic of the system can be described as a series of sequential tunnelling events connecting different many-body eigenstates of the system. For this reason a central role is played in the theory by the spectrum that we calculated in the previous section and by the tunnelling amplitudes between the corresponding many-body energy eigenstates which is the focus of the present one.

Following \cite{leoepjb2008} the 3D electron annihilator in the quantum dot ${\hat{\Psi}}_\sigma(\vec{r})$ can be expressed  in terms of the slow varying 1D operators $\hat{\psi}_{rF\sigma}(x)$ which, in their bosonized form, read:
\begin{equation}\label{eq:psi_x}
\hat{\psi}_{rF\sigma}(x) = \hat{\eta}_{r\sigma}\hat{K}_{rF\sigma}(x)
\rme^{\rmi\hat{\phi}^\dag_{rF\sigma}(x)}
\rme^{\rmi\hat{\phi}_{rF\sigma}(x)}
\end{equation}
where $\hat{\eta}_{r\sigma}$ is the Klein factor which reduces by one the occupation of the branch $r\sigma$, $\hat{K}_{rF\sigma}(x)$ is the operator
\begin{equation}\label{eq:k_x}
\hat{K}_{rF\sigma}(x) = \frac{1}{\sqrt{2L_{\rm d}}} \rme^{\rmi \frac{\pi}{L_{\rm d}}{\rm sgn}(F)(r\hat{N}_{r\sigma} + \Delta)x}
\end{equation}
which essentially adds a phase proportional to the occupation number of the branch $r\sigma$, and $\hat{\phi}_{rF\sigma}(x)$ is the bosonic field associated with the bosonic excitation of the SWCNT.
It is useful to express $\hat{\psi}_{rF\sigma}(x)$ in terms of the position and momentum operators of the plasmonic modes $\hat{X}_n$, and $\hat{P}_n$. After a lengthy but straightforward calculation one obtains:

\begin{equation}\label{eq:psi_plasmon}
\hat{\psi}_{rF\sigma}(x) \propto \hat{\eta}_{r\sigma}\hat{K}_{rF\sigma}(x)
\prod_{n \geq 1}
\rme^{
+\rmi P_n(x)\hat{X}_n
-\rmi X_n(x)\hat{P}_n
}
\end{equation}
where we have introduced the functions
\begin{equation}
\eqalign{
X_n(x) = \sqrt{\frac{2}{ng_{c+}}}
\cos
\left[\frac{n \pi}{L_{\rm d}}
\left(x-x_{\rm d}+\frac{L_{\rm d}}{2}\right)\right],\\
P_n(x) = \sqrt{\frac{2g_{c+}}{n}}{\rm sgn}(Fr)
\sin\left[\frac{n \pi}{L_{\rm d}}\left(x-x_{\rm d}+\frac{L_{\rm d}}{2}\right)\right]
}
\end{equation}
and the proportionality in (\ref{eq:psi_plasmon}) is due the frozen $c-,\,s+$ and $s-$ branches. They only contribute in fact with an overall constant to the tunnelling matrix elements between the low energy eigenstates.

\begin{figure}[h!]
\begin{center}
  \includegraphics[width=0.7\textwidth]{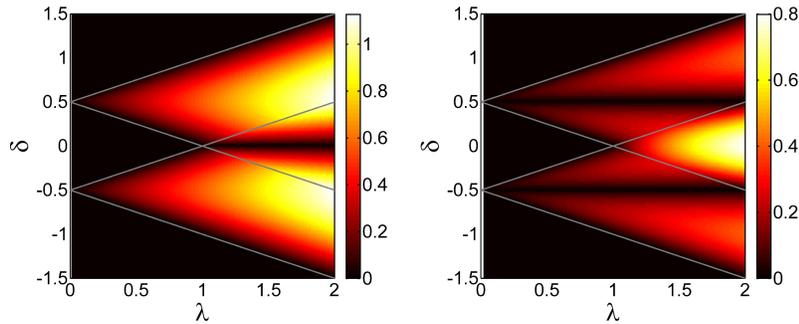}
  \caption{Franck-Condon couplings $|\lambda_l|$ as a function of the length ratio $\lambda = L_{\rm v}/L_{\rm d}$ and relative center position $\delta = (x_{\rm v}-x_{\rm d})/L_{\rm d}$. The coupling for the first and second vibron-plasmon modes are shown in the left and right panel, respectively. The parameters are the same as those reported in figure \ref{fig:DeltaE}. The couplings are calculated for a tunnelling event at the beginning of the dot.}
  \label{fig:FC_lambdadelta}
  \end{center}
\end{figure}

\begin{figure}[h!]
\begin{center}
  \includegraphics[width=0.7\textwidth]{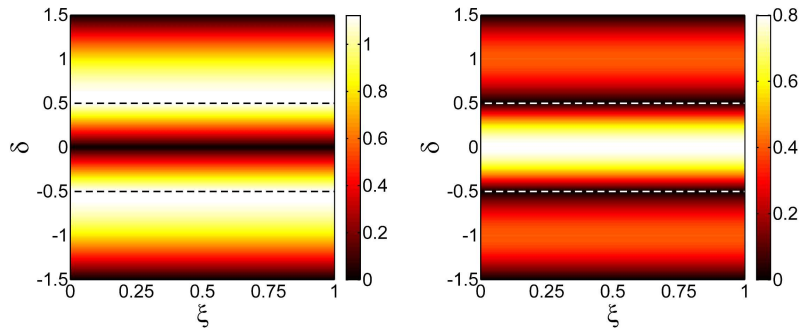}
  \caption{Franck-Condon couplings of the first (left) and second (right) vibron-plasmon mode in the long vibron regime ($\lambda = 2$) plotted against the dimensionless tunnelling point $\xi = x/L_{\rm d}$ and the relative position $\delta$ of the dot and vibron centers. Dashed lines indicate the borders between the B (top), A (center), and D (bottom) regions of the parameters space (see figure \ref{fig:Phase_space}).}
  \label{fig:FC_longvibron}
  \end{center}
\end{figure}

An explicit representation of these low energy eigenstates is readily obtained from (\ref{eq:H_N}) and (\ref{eq:H_sys_pol}). Due to the already mentioned energy scale separation between on one side the vibronic and on the other side the plasmonic and electronic excitations, we can limit ourselves, without loss of generality, to the case $m_{n,j} = 0, j \neq c+$ and obtain

\begin{equation}\label{eq:Eigenstate1}
|\vec{N},\vec{m}\rangle = \rme^{\hat{S}}|\vec{N},\vec{m}\rangle_0
\end{equation}
where
\begin{equation}\label{eq:Eigenstate2}
|\vec{N},\vec{m}\rangle_0 = \prod_l
\frac{(\hat{\xi}_l - \rmi \hat{\pi}_l)^{m_l}}{\sqrt{2m_l!}}|\vec{N},0\rangle_0.
\end{equation}
with $\vec{N} = [N_{c+}, N_{c-}, N_{s+}, N_{s-}]$ being the vector defining the electronic configuration and $\vec{m}$ representing here the occupation numbers of only the lowest vibron-plasmon modes (with an excitation energy lower that $\varepsilon_0$). The low energy eigenstates of a metallic suspended SWCNT are, thus, polaron shifted plasmon-vibron excitations over its electronic ground state $|\vec{N},0\rangle_0$. We are now ready to evaluate the matrix element:

\begin{equation}
\langle \vec{N},\vec{m}|\hat{\psi}_{rF\sigma}(x)|\vec{N}',\vec{m}'\rangle = \,_0\langle \vec{N},\vec{m}|\rme^{-\hat{S}}\hat{\psi}_{rF\sigma}(x)\rme^{+\hat{S}}|\vec{N}',\vec{m}'\rangle_0.
\end{equation}
Since the operator $\hat{S}$ defined in (\ref{eq:S_operator}) commutes with $\hat{K}_{rF\sigma}$:

\begin{equation}\label{eq:psi_shift}
\rme^{-\hat{S}}\hat{\psi}_{rF\sigma}(x)\rme^{+\hat{S}} \propto
\hat{\eta}_{r\sigma} \hat{K}_{rF\sigma}\prod_l
\rme^
{+\rmi \pi_l(x)\hat{\xi}_l
-\rmi \xi_l(x)\hat{\pi}_l}
\end{equation}
where the proportionality accounts for the constant terms deriving by the application of the Baker-Hausdorff theorem and we defined the functions:
\begin{equation}\label{eq:xipi_l}
\eqalign{
\xi_l(x) =& -\frac{\sqrt{2}I}{\varepsilon_l}\sum_{m = 1}^{N_{\rm v}}\sqrt{\frac{\hbar\omega}{\varepsilon_l}}m L_m U_{N_{\rm p}+m,\,l}\\ &+ \sum_{n = 1}^{N_{\rm p}} \sqrt{\frac{2\varepsilon_l}{n^2g_{c+}\hbar\Omega}}U_{nl}
\cos
\left[
\frac{n\pi}{L_{\rm d}}
\left(x-x_{\rm d}+\frac{L_{\rm d}}{2}\right)
\right],\\
\pi_l(x) = &\sum_{n=1}^{N_{\rm p}}\sqrt{\frac{2g_{c+}\hbar\Omega}{\varepsilon_l}}U_{nl}
\sin
\left[
\frac{n\pi}{L_{\rm d}}
\left(x-x_{\rm d}+\frac{L_{\rm d}}{2}\right)
\right].
}
\end{equation}
By means of (\ref{eq:psi_shift}) it is now clear that the tunnelling matrix element factorizes into an electronic component and a product of Franck-Condon factors, one for each plasmon-vibron mode:
\begin{equation}
\langle\vec{N},\vec{m}|\hat{\psi}_{rF\sigma}(x)|\vec{N}',\vec{m}'\rangle
\propto
\langle \vec{N}|\hat{\eta}_{r\sigma} \hat{K}_{rF\sigma}|\vec{N}'\rangle
\prod_l F(m_l,m'_l,\lambda_l)
\end{equation}
where
\begin{equation}
\lambda_l = -\frac{\xi_l - \rmi\pi_l}{\sqrt{2}}
\end{equation}
is the effective coupling between the charge and the plasmon-vibron mode and
\begin{equation}\label{eq:FC_factors}
\eqalign{
F(m,m',\lambda) &=
\left[
\theta(m'-m)\lambda^{m'-m}+
\theta(m-m')(-\lambda^*)^{m-m'}\right]\\
&\times
\sqrt{\frac{m_{\rm min}!}{m_{\rm max}!}}\sum_{i=0}^{m_{\rm min}}
\frac{(-|\lambda|^2)^i}{i!(i+m_{\rm max}-m_{\rm min})!}\frac{m_{\rm max}!}{(m_{\rm min}-i)!}
}
\end{equation}
is the explicit expression of the Franck-Condon factor. The equations (\ref{eq:xipi_l})-(\ref{eq:FC_factors}) together with (\ref{eq:Canonical_2}) for the definition of the transformation $U$ represent the main analytical result of this paper. They are a very general expression of the tunnelling matrix elements between the low energy eigenstates of a suspended SWCNT in presence of multiple plasmon and vibron modes. Special limits of these formulas are already present in the literature \cite{izumida2005, CavalierePRB(R)2010}. Particularly interesting to our point of view is the contribution of the geometrical configuration of the junction, which determines selection rules in the tunnelling processes and in turn the magnitude of the dimensionless electron-vibron Franck-Condon couplings $\lambda_l$. In figure \ref{fig:FC_lambdadelta} we present $|\lambda_l|$ for the first (left) and the second (right) plasmon-vibron modes.
%
%
The values in the figure correspond to a tunnelling matrix element calculated at the beginning of the tube $(x = x_{\rm d}-L_{\rm d}/2)$. By a comparison with the corresponding charge-vibron coupling constant $L_m$ in the lower panels of figure (\ref{fig:KL_functions}) one can argue that $|\lambda_m| \propto |L_m|$. This observation is essentially correct, at least in the A, B and D regions of the parameters space where the energy renormalization of the vibronic modes is negligible and the same holds for the mixing introduced in equation (\ref{eq:Canonical_2}) between the vibronic and plasmonic modes. Consequently, we expect that $|\lambda_l|$ does not depend on the tunnelling point, at least in the long vibron region ($\lambda > 1$) for any geometrical configuration. This result is illustrated in figure \ref{fig:FC_longvibron} where the Franck-Condon couplings for the first and second plasmon-vibron modes are plotted as a function of the dimensionless tunnelling point ($\xi = x/L_{\rm d}$) and relative position of the vibron ($\delta$) for the configuration $L_{\rm v}/L_{\rm d}=2$. Interestingly, the selection rules derived in the previous section for $L_m$ directly apply to the Franck-Condon couplings in the long vibron regime: for example for a symmetric junction ($\delta \approx 0$) only even modes can be excited by a tunnelling event while the odd ones will remain in their ground state. Finally, it is also notable that $\max(|\lambda_l|) \approx 1$ in the long vibron regime even in absence of strong screening ($g_{c+} \approx 0.2$).

\begin{figure}[h!]
\begin{center}
  \includegraphics[width=0.7\textwidth]{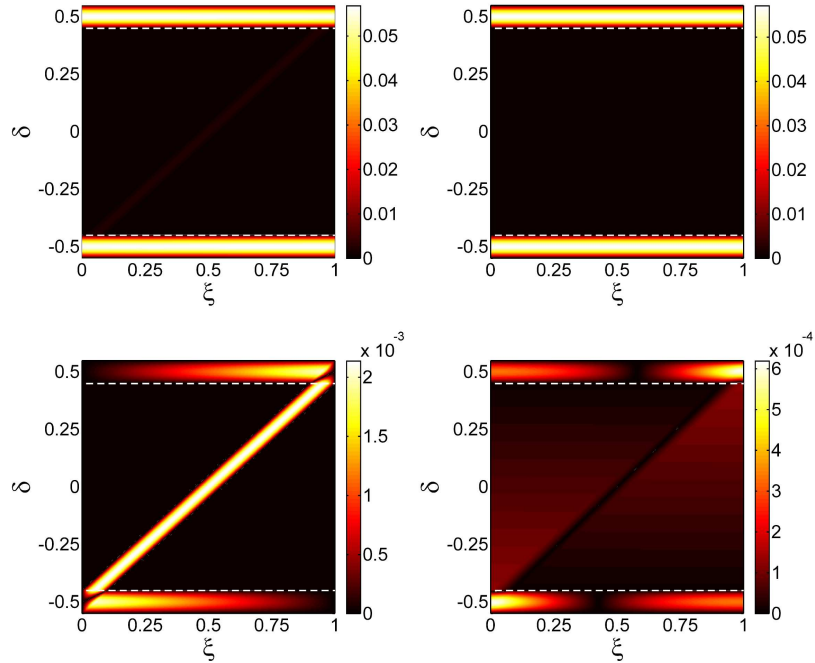}
  \caption{Position dependence of the Franck-Condon coupling of the first vibron-plasmon mode in the short vibron regime ($\lambda = 0.1$). In the upper left panel the full coupling $|\lambda_1|$ is plotted while in the remaining panels its different components: {\it i.e.} in the upper right panel the charge-vibron component -- the first line of $\xi_l$ in (\ref{eq:xipi_l}) --, in the lower left panel the plasmon-vibron component of $\xi_l$, and $\pi_l$ in the lower right panel. The dashed white lines represent in all panels the borders of the C region {\it i.e.} the region in which the vibron is completely inside the dot.}
  \label{fig:FC_shortvibron1}
  \end{center}
\end{figure}

\begin{figure}[h!]
\begin{center}
  \includegraphics[width=0.7\textwidth]{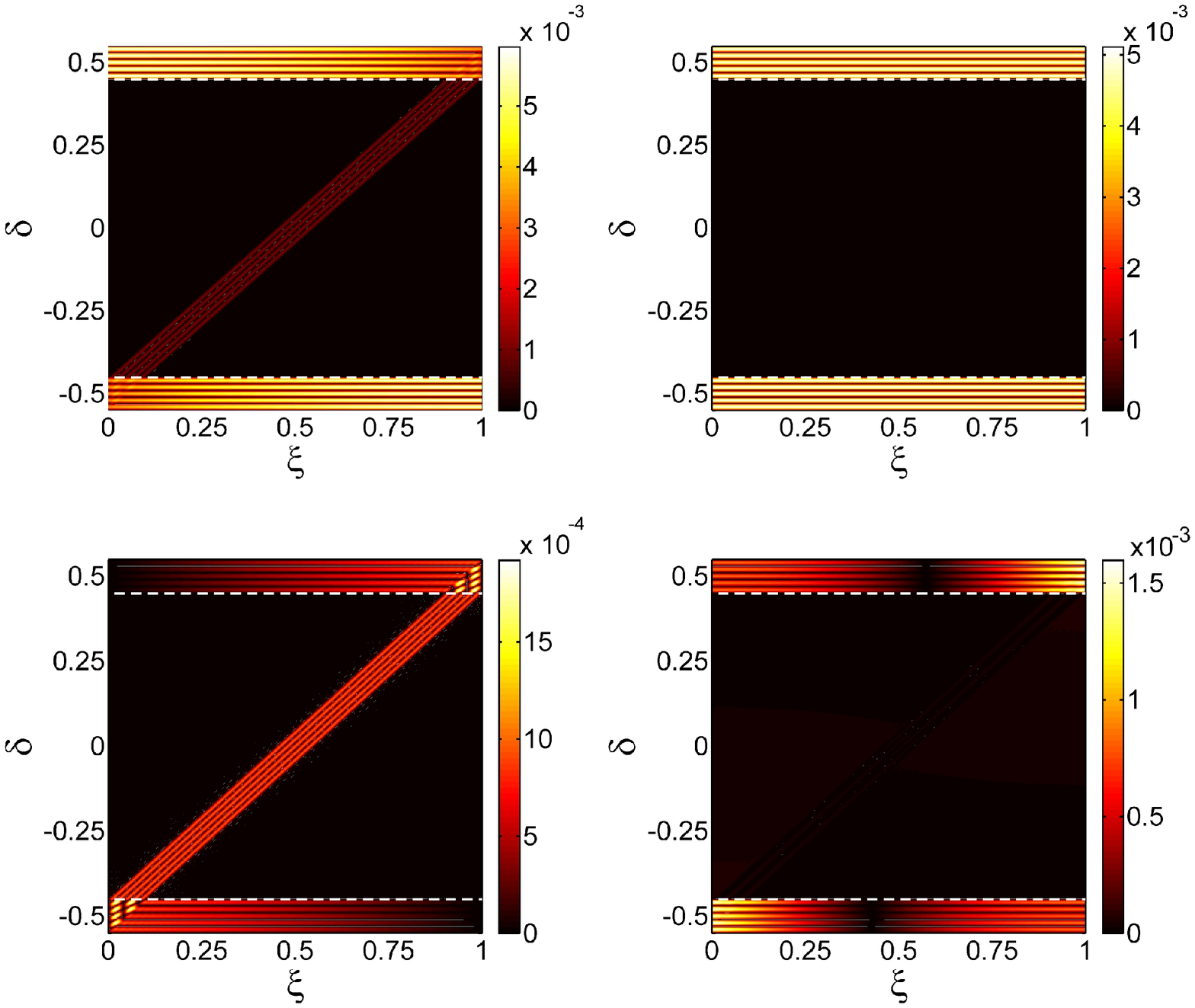}
  \caption{Position dependence of the Franck-Condon coupling of the fifth vibron-plasmon mode in the short vibron regime ($\lambda = 0.1$). In the upper left panel the full coupling $|\lambda_5|$ is plotted. As in figure \ref{fig:FC_shortvibron1} the other panels represent its different components.}
  \label{fig:FC_shortvibron5}
  \end{center}
\end{figure}

A different result characterizes the short vibron limit ($\lambda < 1$). In the C region the charge-vibron coupling vanishes identically due to symmetry considerations. Even if small, the vibron-plasmon mixing becomes there the dominant effect. In figure \ref{fig:FC_shortvibron1} we present the Franck-Condon coupling of the lowest vibron-plasmon mode for the configuration $\lambda = L_{\rm v}/L_{\rm d} = 0.1$. In particular, in the upper left panel we show $|\lambda_1|$ and in the remaining panels its components: {\it i.e.} in the upper right panel the charge-vibron component -- the first line of $\xi_l$ in (\ref{eq:xipi_l}) --, in the lower left panel the plasmon-vibron component of $\xi_l$, and $\pi_l$ in the lower right panel. The dashed white lines represent in all panels the borders of the C region {\it i.e.} the region in which the vibron is completely inside the dot. Outside the C region the charge-vibron coupling is stronger and $\lambda_1$ does not depend on the tunnelling point. Inside the C region, instead, the dominant contribution is given (in the lower left panel) by the plasmon-vibron component of $\xi_l$. The latter follows  the position of the vibron and mimics its shape. The last observation is also confirmed by the lower left panel of figure \ref{fig:FC_shortvibron5} where the corresponding component of the Franck-Condon coupling for the fifth mode is plotted. \hlA{Finally, for higher modes in the short vibron limit, a position dependent Franck-Condon coupling is still appreciable also in the B and D regions (see figure \ref{fig:FC_shortvibron5}).

 The relevance of these results for the tunnelling Hamiltonian and the associated tunnelling rates between the many-body eigenstates depends by their interplay with the spatially dependent transparency $T(\vec{r})$ introduced in the beginning of this section.
 In fact we would expect to detect a position dependent Franck-Condon factor in the tunnelling rates only for the cases illustrated in figures 1A, 1A' and 1B where the vibron extends also beyond the dot region but not for the case in figure 1C. This observation, together with the results presented in figures \ref{fig:FC_longvibron}-\ref{fig:FC_shortvibron5} allows to conclude that the position dependent rates can be observed, among the configurations considered in this paper, only in the asymmetric short vibron one ($\lambda \ll 1,\,\, \delta \approx \pm1/2$), i.e. a configuration of type B (or D), also in agreement with the results presented in \cite{CavalierePRB(R)2010}. In a recent publication \cite{ziani2011} an alternative set-up has been proposed for the direct visualization of the position dependent Franck-Condon couplings in which one of the two metallic electrodes is substituted by the tip of a scanning tunnelling microscope.}

In absence of electron-vibron coupling the frequency of the $n$th stretching mode is an $n$th multiple of the frequency $\omega$ of the fundamental mode. Hence, naturally, there are several energetic degenerate vibronic configurations (involving two or more modes) which may contribute to transport at finite bias. As we just proved, for realistic values of the parameters the softening of the stretching modes introduced by the electron-vibron coupling does not really lift these degeneracies. This fact has profound implications for the transport properties of the system. Interference effects have been in fact predicted even for systems in the Coulomb blockade regime \cite{begemannPRB2008, donariniPRB2010, schultz2010, schultzPRB2009} in presence of quasi-degenerate states.

Technically, this degeneracy determines the method of choice for the description of the dynamics of the system. At low biases, such that only the lowest vibronic mode is excited, a description of the dynamics only in terms of rate equations involving occupation probabilities of the many-body states of the quantum dots is appropriate. Yet, at higher bias, when several vibron modes are excited, a generalized master equation (GME) coupling diagonal (populations) and off-diagonal (coherences) elements of the quantum dot reduced density matrix should be used (see {\it e.g.} (\cite{ begemannPRB2008, schultz2010, schultzPRB2009, braigPRB2005, braunPRB2004,  wunschPRB2005, harbolaPRB2006, mayrhoferEPJB2007, kollerNJP2007, hornberger2008}).

The sensitive dependence of the tunnelling matrix elements on the mode number for a given geometry of the system also suggests the existence of symmetrically coupled slow channels as the ones described in \cite{YarPRB2011} and consequently of similar NDC effects in the stability diagrams of a suspended  SWCNT junction.

\section{Conclusions}

In this paper we analyzed the spectrum and the effective Franck-Condon couplings of a suspended SWCNT quantum dot including many vibronic modes as well as different dot-vibron geometrical configurations. We described the long-wavelength acoustic-vibrons within an elastic continuum model and the electron-vibron interaction in terms of a deformation potential. In particular, we studied how the electromechanical properties depend on the relative size $\lambda$ and position $\delta$ of the vibron with respect to the dot.

Specifically, within the framework of a Tomonaga-Luttinger liquid description of the SWCNT, we derived an effective low energy Hamiltonian where the electron-vibron coupling is separated into a plasmon-vibron and a charge-vibron component proportional to different coupling constants ($K_{nm}$ and $L_m$, respectively).

The system was diagonalized via a series of canonical transformations with an intuitive geometrical interpretation which reduce the low energy description of the suspended SWCNT to a set of displaced plasmon-vibron excitations. Consequently, the tunnelling matrix elements between the many-body eigenstates of the system are the product of Franck-Condon factors, one for each plasmon-vibron mode, of which  we gave an analytical expression.

The analysis of the coupling constants $K_{nm}$ and $L_m$ and of the Franck-Condon couplings $\lambda_l$ on the entire geometrical parameters space allowed us to identify different regimes.

In the short symmetric vibron regime ($\lambda < 1$, $\delta \approx 0$) the charge-vibron component vanishes and the Franck-Condon couplings are extremely small ($|\lambda_m|\approx10^{-3}$) due to the energy scale separation between the plasmonic and vibronic modes ($\Omega/\omega \gg 1$ and $\hbar\Omega/I \gg 1 $) that hinder the plasmon-vibron mixing. The Franck-Condon coupling is position dependent and is located around the position of the vibron.

In the long vibron regime ($\lambda \gg 1$) the charge-vibron coupling dominates the scenario giving substantially larger Franck-Condon couplings ($|\lambda_m| \approx 1$) and independent of the position as in the simple Anderson-Holstein model. The Franck-Condon couplings are strongly dependent on the relative position of the vibron and the dot, leading to selections rules: for example only even vibron-plasmon modes can be excited by electron tunnelling in a symmetric ($\delta = 0$) long vibron junction (see figure \ref{fig:FC_lambdadelta}).

In the asymmetric short vibron regime ($\lambda < 1$, $\delta \approx \pm 1/2$) the charge-vibron and plasmon-vibron contribution are of the same order and correspondingly one can distinguish (at least in the higher modes, see figure \ref{fig:FC_shortvibron5}) the position dependent contribution due the plasmon-vibron mixing superimposed to the uniform polaron shift typical of the charge-vibron component of the coupling. In absence of screening (dimensionless electron-electron interaction strength $g_{c+} = 0.2$), though, the absolute value of the Franck-Condon coupling remains negligibly small compared to the one estimated from the experiments \cite{sapmaz2006,huettel2009,{leturcq09}}. Reasonable values have been obtained in this regime by \cite{CavalierePRB(R)2010} by assuming a very strong screening ($g_{c+} \approx 1$) that essentially removes the energy scale separation between the plasmon modes and the much shorter vibron mode.

Finally, for reasonable values of the nanotube parameters the spectrum of the nanotube is only slightly modified by the electron-vibron coupling thus preserving the high degeneracy of the different vibronic configurations. This, in combination with the sensitive dependence of the tunnelling matrix elements on the mode number and on the geometry of the system, also suggests the existence of symmetrically coupled slow channels as the ones described in \cite{YarPRB2011} and consequently of similar NDC effects in the stability diagrams of a suspended nanotube junction.

\ack Support of the DFG under the programs SFB 689 and GRK 1570 is acknowledged. Abdullah Yar also acknowledges the support of Kohat University of Science \& Technology, Kohat-26000, Khyber Pakhtunkhwa, Pakistan under the Human Resource Development Program.

\end{document}